\documentclass[12pt,preprint,landscape]{aastex}

\newcommand{\kms}{km s$^{-1}$}
\newcommand{\et}{\it et al.\rm}

\newcommand{\mjb}{mJy beam$^{-1}$}

\newcommand{\HI}{\ion{H}{1}}
\newcommand{\CI}{\ion{C}{1}}
\newcommand\Blos{$B_{los}$}
\newcommand\muG{$\mu$G}
\newcommand{\src}{\objectname[3C]{3C~138}}
\newcommand{\cc}{cm$^{-3}$}

\begin{document}

\title{Spatial and Temporal Variations in Small-Scale Galactic \HI\/ 
Structure Toward 3C~138}

 \author{C. L. Brogan\altaffilmark{1,2}, B. A. Zauderer\altaffilmark{3},
  T. J. Lazio\altaffilmark{4},
  W. M. Goss\altaffilmark{5}, C. G. DePree\altaffilmark{6}, and 
  M. D. Faison\altaffilmark{7}}
                                                                       
\altaffiltext{1}{Institute for Astronomy, 640 North A'ohoku Place, Hilo, 
HI 96720}

\altaffiltext{2}{JCMT Fellow}

\altaffiltext{3}{Department of Astronomy, University of Maryland, College 
Park, MD 20740}

\altaffiltext{4}{Naval Research Laboratory, Remote Sensing Division, 
Code 7213, 4555 Overlook Avenue SW, Washington, DC 20375-5351}

\altaffiltext{5}{National Radio Astronomy Observatory, P. O. Box O, 1003 
Lopezville Road, Socorro, NM 87801}

\altaffiltext{6}{Department of Physics and Astronomy, Agnes Scott
  College, 141 E. College Ave., Decatur, GA 30030}
  
\altaffiltext{7}{Reed College, Dept. of Physics, 3203 SE
  Woodstock Blvd., Portland, OR 97202}

\begin{abstract}

Several recent studies of Galactic \HI\/ absorption toward background
quasars and pulsars have provided evidence that there are opacity
changes in the neutral Galactic interstellar medium on size scales as
small as a few AU.  The nature of these opacity variations has
remained a matter of debate, but could reflect a variety of physical
processes, including changes in the \HI\/ spin temperature or gas
density.  We present three epochs of VLBA observations of Galactic
\HI\/ absorption toward the quasar 3C~138 with resolutions of 20 mas
($\sim 10$ AU).  This analysis includes VLBA data from observations in
1999 and 2002 along with a reexamination of the 1995 VLBA data,
reported by \citet{fgdt98}.  Improved data reduction and imaging
techniques have led to an order of magnitude improvement in
sensitivity compared to previous work.  With these new data we confirm
the previously detected milliarcsecond scale spatial variations in the
\HI\/ opacity at the level of $\Delta\tau_{max} =0.50\pm0.05$.  The
typical size scale of the optical depth variations is $\sim 50$ mas or
25 AU.  In addition, for the first time we see clear evidence for
temporal variations in the \HI\/ opacity over the seven year time span
of our three epochs of data.  We also attempted to detect the magnetic
field strength in the \HI\/ gas using the Zeeman effect.  From this
analysis we have been able to place a 3 $\sigma$~upper limit on the
magnetic field strength per pixel of $\sim 45~\mu$G.  We have also
been able to calculate for the first time the plane of sky covering
fraction of the small scale \HI\/ gas of $\sim 10\%$. This small
covering fraction suggests that the filling factor of such gas is
quite low in agreement with recent optical observations. 
We also find that the line widths of the milliarcsecond sizescale
\HI\/ features are comparable to those determined from previous single
dish measurements toward 3C~138, suggesting that the opacity
variations cannot be due to changes in the \HI\/ spin temperature.
From these results we favor a density enhancement interpretation for
the small scale \HI\/ structures, although these enhancements appear
to be of short duration and are unlikely to be in equilibrium.

\end{abstract}


\section{Introduction}

A variety of absorption line studies at wavelengths from radio to
optical over the past three decades have found AU-scale optical depth
variations in the interstellar medium (ISM). The first 21 cm evidence
for changes in \HI\/ opacity over small scales was provided by
\citet{dwr76} using a single baseline VLBI observation of 3C~147.
These results were confirmed by \citet{dgrbkm89} and the first images
of the small scale \HI\/ in the direction of 3C~138 and 3C~147 were
made by \citet{ddg96} using the MERLIN array.  These results were
extended to higher resolution by \citet{fgdt98} and \citet{fg01}
using the newly completed VLBA with the phased VLA to image \HI\/ opacities
with an angular resolution of $\sim$20 mas or $\sim$10 AU (see below).
Significant variations were detected in the direction of 3C~138 and
3C~147 while no significant variations in \HI\/ opacity were found in
the direction of five other compact radio sources. Thus far small scale 
\HI\/ absorption studies toward quasars have lacked temporal data on such 
variations.

Many pulsars have been observed to move with transverse speeds greater
than 100 \kms\/.  Thus, \HI\/ observations spaced over a several year
period will sample different lines of sight.  The pulsar data have
presented conflicting results.  \citet{fwcm94} used the Arecibo
telescope over a period of 1.7 years and found \HI\/ opacity changes
in six of seven objects.  \citet{dmra92} observed two pulsars and find
marginal evidence for \HI\/ opacity in the direction of one pulsar
(PSR 1557-50) over a four year period.  \citet{jkww03} have
re-observed PSR 1557-50 and find significant changes in the \HI\/
opacity over a 2.5 year period.  \citet{swhdg03} have re-observed
three of the pulsars in the sample of \citet{fwcm94}, using the
Arecibo telescope over time scales of 9$-$10 years, or spatial scales
of 10$-$20 and 70$-$85 AU.  No significant \HI\/ opacity changes were
observed and \citet{swhdg03} suggest that AU sized structures in the
ISM may be quite rare.  As we will show in \S 4.2, the one-dimensional
nature of the pulsar sampling may mean that small scale structures in
the ISM with low filling factors could be missed in many cases.

Optical fine structure absorption lines from ionized and neutral atoms
show pronounced spatial variations toward binary stars with
separations of a few 100s of AU \citep[see for
example][]{Jenkins04}. In a few recent studies, temporal variations
have also been observed over timescales of a few years
\citep{ml99,lmb00}.  Observations of the optical lines of interstellar
molecules such as those of CH by \citet{rbfp03}, have also shown
variations as a function of time. Additionally, radio observations of
interstellar H$_{2}$CO using the VLA have shown time variations
\citep{mmb93,mm95} consistent with the optical observations of
CH. However, \citet{lmwb98} point out that the variations of the
interstellar lines in the direction of the $\mu$ Crucis binary system
are only prominent in the rarer ionization states.  For example, the
predominant ionization state \ion{Zn}{2} shows little variation, while
the \ion{Na}{1} D line from rarer neutral sodium shows prominent
variations. This discrepancy suggests that the observed variations are
due to small differences in the physical properties of the ISM such as
temperature, or ionization state, rather than being the result of
distinct physical entities.
 
A number of theoretical studies of the nature of small scale Galactic
structure have been made in the last few decades in order to
understand this puzzling phenomenon.  If the changes in \HI\/ opacity
are interpreted in terms of roughly spherical entities with a constant
spin temperature ($\tau\propto N_H/T_{spin}$ and $N_H = n_H*l$, where
$l$ is the diameter of the feature), then small ($\sim$ 10 AU) dense
($n_H\sim 10{^5}- 10{^6}$ cm$^{-3}$) \HI\/ features are implied.
Numerous authors such as \citet{dgrbkm89}, have discussed the implied
overpressure and short time scale problems that such structures would
have.  Several attempts have been made in recent years in order to
alleviate these problems.  \citet{h97} proposed that the TSAS
(tiny-scale atomic structure) might be explained by a combination of
lower spin temperatures and geometry; for example, the \HI\/ features
could be elongated cylinders or sheets with the long axis parallel to
the line of sight, thus increasing the path length along the line of 
sight and reducing the required density.
 
In contrast, \citet{d00} has used one of the power spectra derived by
\citet{ddg00} for the size scales of Galactic \HI\/ to suggest that
the opacity variations are not physical entities, but are simply a
manifestation of random fluctuations contributed by the entire range
of spatial scales in the "red" (i.e. noise that increases with larger
spatial frequencies) \HI\/ power spectrum.  However, the VLA \HI\/
data toward CasA used to construct the \citet{ddg00} power spectrum
has a resolution of $7\arcsec$, and thus probes linear dimensions in
the range 0.07 to 3 pc, compared to the $\sim 10$ AU size scales
($5\times 10^{-5}$ pc) probed by the VLBA observations.  Despite the
large difference between the physical size scales that were actually
measured, this extrapolation of the CasA data from a few thousand AU
to scales of tens of AU predicts fluctuations at the level of
$\lesssim$ 0.1 in opacity, which is consistent with many, but not all,
of the VLBA \HI\/ results of \citet{fgdt98} and \citet{fg01}.

The extragalactic source 3C~138 was selected to continue investigation
into small scale structure in neutral \HI\/ opacities for a variety of
reasons. The size and brightness of 3C~138 makes it a favorable
candidate for an \HI\/ absorption study. In addition, among the seven
sources examined by \citet{fgdt98} and \citet{fg01}, 3C~138 shows the
most significant spatial variations in optical depth.  Thus it is the
most promising candidate to re-examine both the level of spatial
variations, as well to probe for the first time \HI\/ temporal
variations. Additionally, the high signal to noise absorption spectra
afforded by such a strong continuum source can be used to measure the
line of sight magnetic field strength in the small scale neutral gas
via the Zeeman effect.  An upper limit on distance to the \HI\/ gas is
500 pc, based on the Galactic latitude of the source ({\it b} =
$-11.34^{\circ}$) and the assumption that the cold atomic \HI\/ layer
is no thicker than 100 pc \citep{fgdt98}.
 
In this paper we present new VLBA\footnote{The National Radio
Astronomy Observatory (NRAO) is a facility of the National Science
Foundation operated under a cooperative agreement by Associated
Universities, Inc.} plus the phased VLA images of the \HI\/ absorption
in the direction of 3C~138 for two epochs taken in 1999 and 2002.
These data are compared to the first epoch of 3C~138 VLBA \HI\/ data
taken in 1995 and presented in \citet{fgdt98}.  The organization of
the paper is as follows: A description of the new observations and
their data reduction are presented in \S 2.  The \HI\/ opacity results
are presented in \S 3 including the time variations observed in a
comparison of the three epochs over a seven year period.  A discussion
of our results is presented in \S 4.  Our conclusions follow in \S 5.

\section{OBSERVATIONS AND DATA REDUCTION}

We have observed the Galactic \HI\/ absorption (near $\sim 0$ \kms\/) 
toward the extragalactic source 3C~138 for three separate epochs
taken in 1995 (Epoch I), 1999 (Epoch II), and 2002 (Epoch III). The
results from Epoch I have been previously published in \citet{fgdt98},
while preliminary results from Epoch II were presented in
\citet{zfgbd02}. Table 1 summarizes the basic observing parameters for
all three epochs.  For each epoch the data were obtained using the ten
antennas of the VLBA combined with the 27 antennas of the phased
VLA. The GBT and Arecibo were also included in the 2002 data, but were
later removed due to technical problems at both telescopes during the
observation.  Each observation lasted for 12 hours (including time spent on
calibration sources).  The proximity of the VLA to the VLBA antenna at
Pie Town, New Mexico significantly increases our sensitivity to large
scale structures.  Note that given the different rise times of 3C~138
at the different VLBA stations, the more distant (from the VLA)
antennas: St. Croix, Hancock, and Mauna Kea were not on source
throughout the 12 hours of the observations.  For example, the
St. Croix antenna was typically only included for $\sim$~60\% of the
total observing period, and was lost completely during the Epoch I
observations. Henceforth, reference to ``VLBA'' will refer to the
combined VLBA plus phased VLA. The shortest baseline included in 
these data is $\sim 60$ km (PT-phased VLA) so that structures 
larger than $\sim 700$ mas are resolved out.

In order to obtain high spectral resolution at the velocities of
Galactic \HI\/ absorption ($\sim 0$ \kms\/ ) and good continuum
sensitivity uncontaminated by \HI\/ absorption for calibration
purposes, data were obtained in four separate IFs with center
velocities of $-100$, 5, 105, 200 \kms\/  for Epoch I and $-180, 5,
180, 360$ \kms\/ for Epochs II and III. Both right and left circular
polarizations were recorded for each IF. A summary of the different
bandwidths and spectral resolutions used in each epoch is given in
Table 1.

The data reduction and imaging for all three epochs were performed using
the NRAO AIPS software package. The initial amplitude calibration was
applied in the usual manner using the calibration tables supplied by
the VLBA.  The delays were measured by fringe fitting on the strong
continuum source 3C~84. Bandpass calibration was also determined from
observations of 3C~84. Corrections for delay tracking were then
applied using the task CVEL.

A number of steps were subsequently taken to ensure uniformity in the
3C~138 data reduction from epoch to epoch as small changes in the
continuum or line images resulting from differences in calibration
could mimic optical depth changes. First, the published 3C~138
continuum image from the 1995 Epoch I data \citep[see][]{fgdt98} was
used as the input model in FRING to solve for improved delays and the
rates for the 2002 Epoch III 3C~138 data. For this step, an average of
all four IFs was used in the FRING solution for maximum
signal-to-noise.  After FRING, several iterations of phase
self-calibration were carried out on the 2002 4IF 3C~138 ``continuum''
dataset. These corrections were applied and IF 3 (with Galactic \HI\/
absorption) was split off. A final 2002 continuum image was then
created from the line-free channels of IF 3, which then underwent a
final round of amplitude and phase self-calibration. These corrections
were also applied to the IF 3 continuum subtracted line data. After
this procedure, the 2002 continuum image was clearly superior to the
\citet{fgdt98} 1995 image. Thus, the 2002 continuum image was used as
the initial model for both the 3C~138 4IF FRING and the 1st round of
4IF phase-only self calibration for the reduction of the 1999 Epoch II
data and a re-reduction of the 1995 Epoch I data.

Beyond ensuring a consistent continuum model, this procedure also
ensures that the three epochs will have self-consistent positions, as
absolute position information is lost when self-calibration is
employed.  During the imaging process the data were essentially
naturally weighted with small adjustments to the robustness parameter
in order to create images with an east-west resolution close to 20
mas. This was necessary to account for small differences in flagging
between the three epochs. The resulting clean beams (which are quite
similar) are listed in Table 1 for reference. Both the continuum
images and continuum subtracted line cubes from all three epochs were
then convolved to exactly 20 mas in both dimensions. This convolution 
should minimize the effects of any remaining differences in the 
$u-v$ coverage of the three datasets. 

The data were Hanning smoothed during the imaging process, so while
the channel separation is 0.41 \kms\/, the velocity resolution is 0.82
\kms\/.  Each channel in the line datasets were cleaned down to the
same final flux level of 30 \mjb\/. The final rms noise
characteristics of the three epochs are listed in Table 1. We have
achieved a significant improvement (a factor of $\sim 5$) in the rms
noise of the Epoch I data over that presented in \citet{fgdt98},
primarily due to improvements in the imaging and cleaning software, as
well as, the iterative continuum modeling procedure described above.

Optical depth cubes were calculated from the IF 3 continuum images and
continuum subtracted line cubes for each epoch using the relation
$\tau = -$ln(${1+{{T_L}\over{T_C}}}$), where T$_L$ is negative. Before
calculating the optical depths, the continuum images were
conservatively masked wherever the flux density is below 60 \mjb\/
($10\%$ of the peak value).  At 500 pc (see \S 1), 20 mas $=\sim 10$
AU. Note that since the \HI\/ gas is likely outside of the local
bubble, a minimum distance is $\sim 100$ pc, implying 20 mas $=\sim 2$
AU. We will assume the far distance and a linear scale of 10 AU in the
remainder of the paper, but the fact that some of the \HI\/ gas likely
lies at smaller distances (and thus a smaller size scale is probed)
should be borne in mind.

\section{RESULTS}

\subsection{21 cm Continuum}

The 21 cm continuum image of 3C~138 from the Epoch III (2002) data is
shown in Figure 1. The resolution of this image is 20 mas and the rms
noise is 0.95 \mjb\/. The brighter parts of the continuum morphology
are in fair agreement with that presented in \citet{fgdt98}, given
that the current image has four times better S/N. The low surface
brightness 21 cm (1.42 GHz) continuum morphology of Figure 1 is in
excellent agreement with the 15 $\times$ 5~mas resolution, 1.7~GHz
VLBI image presented by \citet{Cotton97}. Based on \citet{Cotton97}
the component just to the SW of image center is the core, while the
linear structure to the northeast is the approaching part of the
jet. The low surface brightness component in the SW corner of the
image is the receding jet.  These authors also review previous
observations of 3C~138 and find that the source as a whole is not
significantly variable.  \citet[also see][]{Cotton2003} used high
resolution multi-epoch VLBA data at 5GHz to demonstrate that the core
components change by a few tens of \mjb\/ on timescales of $\sim 2$
years (note that the multiple core components detected by Cotton are
unresolved at 20 mas resolution).  

In order to test for variability in our 21 cm VLBA data, we
constructed difference maps between the continuum images of each
epoch. These maps show that indeed the ``core'' changes from epoch to
epoch by $\sim 50$ \mjb\/, while the rest of the source is constant to
within $\sim 10$ \mjb\/. Note that variability will not impact our
optical depth calculations since: (1) the continuum appropriate for
each epoch was used in the optical depth calculations; (2) no
amplitude self-calibration was transfered from epoch to epoch; and (3)
the intrinsic continuum morphology at 20 mas resolution does not
change from epoch to epoch \citep[according to][positional changes in
the core components over eight years only amounts to $\la 1.7$ mas,
and these components are themselves not resolved by our 20 mas
resolution data]{Cotton2003}.

It is notable that our continuum peak flux densities for all three
epochs of $\sim 605$ \mjb\/ (see Table 1) are about $30\%$ lower than
that reported by \citet{fgdt98} (the images have the same resolution).
The integrated flux of our data (including the re-reduction of the
Epoch I data) is also low: $\sim 6.5$ Jy compared to 8.5 Jy reported
in \citet{fgdt98}. It is most likely that the previous data reduction
is in error, given that the VLA 1.4 GHz flux density of 3C~138 is
$\sim 8.5$ Jy (VLA calibrator manual), and it is very unlikely that a
VLBA image could recover so much of the flux of such an extended
source. Indeed, \citet{Cotton97} also recover $\sim 70\%$ or (5.4 Jy)
of the expected VLA flux in their 1.7 GHz VLBA image. We suspect that
the use of a point source model in FRING during the earlier data
reduction \citep{fgdt98} coupled with details of the self-calibration
are to blame for these discrepancies. We are confident that the
current method of data reduction is valid, and in any case this
difference will not affect the optical depths presented below since
the continuum was extracted from the line free parts of the \HI\/
data, and all three of our epochs (including re-reduction of 1995
data) agree within a few percent.

\subsection{\HI\/ Optical Depths}

Figure 2 shows the average Epoch III (2002) optical depth profile along 
with the difference between the average Epoch III and Epoch II spectra. It
is clear from this figure that the \HI\/ spectrum is rich and complex
toward 3C~138 despite its high latitude ($b=11.3\arcdeg$).  Overall, 
the average spectra are quite similar with significant deviations 
only occurring in the narrow velocity component near 6 \kms\/ and 
in a broader component near $-7$ \kms\/.

\citet{HTI03} recently completed the ``Millennium Arecibo 21 cm
Absorption Line Survey'' toward 79 continuum sources, including
3C~138.  As part of their study these authors have made detailed
Gaussian fits to the Arecibo \HI\/ data, being careful to separate
Galactic emission from absorption.  Since 3C~138 is significantly
smaller than the Arecibo beam at 21cm ($\sim 3\arcmin$), the effective
resolution is equal to the size of 3C~138. Thus comparison of our 20
mas resolution data to the Arecibo spectrum provides an interesting
check on whether or not the physical conditions of the gas change on
scales of order $\sim 800$ mas \citep[see for example MERLIN image
in][]{ddg96}. For example, a change in the spin temperature should be
apparent in the line widths.

To first order, the Heiles \& Troland Arecibo absorption spectrum is
quite similar to our VLBA spectra. Therefore, we have used their
results as the initial guesses for our own Gaussian analysis. The
center velocities were initially held fixed at the \citet{HTI03}
values, and then varied slightly to improve the residuals. The
maximum residual in our optical depth fit occurs near $V_{LSR}\sim +7$
\kms\/ and has a value of 0.04.  Table~2 shows the results of our
analysis for the \HI\/ spectra at the continuum peak of the Epoch III
data along with the Arecibo results for comparison \citep{HTI03}. A
Gaussian analysis for spectra of this complexity is unlikely to be
unique, although, our results are in very good agreement with those
obtained by Heiles \& Troland for the four strongest
components. However, although \citet{HTI03} were able to adequately fit
their spectrum with 6 Gaussians, a seventh component at $\sim -7$
\kms\/ was needed to fit the VLBA spectrum. Interestingly, this is
also one of the two regions where there is a significant deviation
between the average Epoch III and Epoch II VLBA spectra (see
Fig. 2). The Arecibo data were taken closest in time to the Epoch II
data, and the $\sim -7$ \kms\/ was indeed less noticeable
then. Although we have included it in our fit in order to directly
compare with the Heiles \& Troland results, the reality of the broad,
weak component near $\sim +1.9$ \kms\/ is rather dubious.

Overall, the agreement between the VLBA data and Arecibo is quite
impressive in center velocity, line width, and $\tau$.  Although we
only present Gaussian fitting results from the 2002 continuum peak,
neither the center velocity or line width changes significantly as a
function of position or epoch. That is, only the optical depth varies
significantly from position to position and across epochs (these
variations are described in more detail in \S3.4).

Using the average \HI\/ spin temperature derived by \citet{HTI03} for
the 3C~138 absorption components of $\sim 50$ K, we find that the
average 2002 VLBA total \HI\/ column density is $\sim~6\times 10^{20}$
cm$^{-2}$. For this computation we have ignored the broad $+1.9$
\kms\/ component mentioned above (also see Table 2) since its reality
seems doubtful.  For comparison the \HI\/ column density observed in
emission in the Arecibo data is $\sim~9.1\times 10^{20}$ cm$^{-2}$.
Interestingly, given the normal assumption that $N_{TOT}\sim 2\times
10^{21}A_v$ and using $A_v\sim 1$ toward 3C~138 \citep[NED
database;][]{Schlegel98} it is clear that much of the total hydrogen
column toward 3C~138 is atomic since the combined warm plus cold \HI\/
column density can account for most of the total hydrogen column. Indeed, 
since 3C~138 is optically visible \citep{Spinrad85}, $A_v\sim 1$ is likely
an upper limit to the extinction strengthening this conclusion.

\subsection{Zeeman Magnetic Field Limits}

From the high S/N 1999 and 2002 \HI\/ data obtained in this study, we
have also attempted to measure the Zeeman effect. That is, Stokes V =
RCP-LCP; Stokes I= RCP+LCP and in the presence of a magnetic field $V=
ZB_{\mathrm{los}}dI/2d\nu$, where Z=2.8 Hz \muG\/$^{-1}$, and \Blos\/
is the line of sight field strength.  After careful inspection of the
Stokes V cubes for both epochs, we do not find any convincing Zeeman
signatures that meet both of the following criteria: (1) Stokes V
larger than $5\sigma$ simultaneously in $both$ epochs; and (2) Stokes
V more than $5\sigma$ over more than one beam width. The Stokes V
spectral line rms noise levels are the same as those given in Table 1
for Stokes I.  

Near the continuum peak for the component with highest S/N at $\sim 6$
\kms\/, we find $3\sigma$ upper limits for both epochs to \Blos\/ of
$\sim 45$ \muG\/ per pixel, with similar upper limits near the core
(see Fig. 1). This upper limit is similar to the $2\sigma$ limit
reported by \citet{fgdt98}, but on a pixel by pixel basis rather than
for an averaged profile. If instead we use profiles that have been
averaged for pixels with absorption deeper than $-0.05$ \mjb\/, we
find an average $3\sigma$ upper limit of 20 \muG\/.  Using the
``Millennium Arecibo 21 cm Absorption Line Survey'' data described in
$\S 3.3$, \citet{HT04} find \Blos\/$=+5.6\pm 1.0$ \muG\/ for the $\sim
6.4$ \kms\/ component, consistent with our averaged upper limit.

A number of recent VLA and VLBA Zeeman results suggest that the
observed magnetic field strength often increases with higher
resolution due to blending of spectral components at poorer resolution
and beam dilution \citep[i.e. higher fields averaged together with
lower field regions or oppositely directed fields averaged within the
beam;][]{Brogan2001,Sarma2001}. The results presented here suggest
that the Arecibo magnetic field results are not strongly affected
by spectral blending (as already implied by the close agreement
between the spectral profiles presented in \S3.2) or beam dilution
effects. Speculation on the significance of these low magnetic 
field detections and upper limits is provided in \S4.3.3.

\subsection{Small Scale Structure}

For the remaining analysis of the small scale \HI\/ structure, we
focus our attention on the 0.9, 6.2, and 9.1 \kms\/ velocity channels,
since they are representative of the four strongest velocity features 
(see Fig. 2 and Table~2).

A position-velocity image for the Epoch III (2002) data along the
cross-cut labeled (a) in Figure 1 is shown in Figure 3. It is clear
from this image that there are discrete regions of higher and lower
optical depth at constant velocities.  Optical depth velocity channel
maps are presented in Figure 4 for three different velocities: 0.9,
6.2, and 9.1 \kms\/ for all three epochs. {\em The average value of
the 2002 optical depth (for each channel) has been subtracted from
the channel maps displayed in Figure 4}.  The difference in regions of
higher and lower optical depth from epoch to epoch in these images is
striking. To aid in assessing the significance of the optical depth
variations presented in Figures 3 and 4, Table 3 lists the average
$\tau$, along with the minimum, maximum, and mean values of the
optical depth $error$ for each epoch for the 0.9, 6.2, and 9.1 \kms\/
channels. As with any optical depth map, the uncertainty in the
optical depths shown in these figures is a non-linear function that is
inversely proportional to the continuum brightness and proportional to
the depth of the absorption. The minimum S/N ratio of the optical
depths displayed in Figures 3 and 4 is 7 (which occurs near edges of
the continuum), while near the continuum peak the S/N is more than
100.

Figure 5 shows an example of both spatial and temporal \HI\/ opacity
variations across 3C~138 at 0.9 and 6.2 \kms\/, along the position
angle indicated by the dashed line in Fig. 1 marked (a). The width of
the shaded region indicates the $\pm 1\sigma$ uncertainty in the
optical depth measurements.  A number of \HI\/ optical depth
``clumps'' vary from the mean by more than $5\sigma$. The typical
linear distance between significant variations is $\sim 50$~mas.  The
temporal variations between the three epochs are also striking along
this cross cut.  The variations along cross-cut (a) are particularly
significant since it avoids the edges of the continuum where the
optical depth uncertainties are greatest.  For comparison, Figure 6
shows a cross cut along the position angle marked (b) in Figure
1. Clearly in this direction there are no significant spatial or
temporal variations.

Since the weighting, $u-v$ coverage, and self-calibration for the line
and continuum data for a given epoch are the same, the enhancements of
the optical depths within a a given epoch (above the noise threshold)
are almost certainly real. For example, if the line data were actually
spatially smooth and the optical depth enhancements within a given
epoch were dominated by subtle differences between the cleaning of the
line and continuum images for the same epoch, the enhancements should
look the same at each velocity within a given epoch and they do not
(note that each channel of the line cubes was cleaned to the same flux
level). 

The possibility for ``technical issues'' (i.e.  small changes in the
$u-v$ coverage, flux, or imaging) to affect the appearance of the
optical depth maps from epoch to epoch are greater (although the data
reduction steps described in \S2 should minimize these
effects). However, if small changes in the continuum morphology from
epoch to epoch were responsible for the temporal changes, the
variations with time should mimic the continuum difference maps
described in \S2 and they do not.  As a specific example, there is a
50 \mjb\/ difference between the continuum flux density at the peak of
the `core' component (see Fig. 1) due to intrinsic variability
\citep{Cotton2003} between the 1999 and 2002 data, yet the optical
depth channel maps shown in Figure 4 for all three velocities at the
location of the core are very similar between these two epochs (to
within the stated $\sigma(\tau)$ noise levels).  Also, apart from the
core component, the 2002 continuum is consistently higher than the
1999 continuum (although the difference is not consistent with a
simple zero point offset).  However, for all three velocity
components, the optical depths are not at all consistent with this trend
(i.e. 1999 $\tau$ consistently higher than 2002; Figs. 4 and 5).
Indeed, difference maps of the 2002 and 1999 {\em optical depths} look
nothing like the continuum difference maps, and are instead very
well correlated with the spectral line difference maps. 

These comparisons demonstrate that the line to continuum ratios and
hence optical depths are insensitive to changes in the continuum, as
expected (i.e. the line changes commensurately, except where there is
a real enhancement).  Assessment of the effect of small changes in the
$u-v$ coverage or imaging on the line data {\em independent from those
apparent in the continuum images} from epoch to epoch is more
difficult. However, there is no reason to believe that these effects
would be {\em greater} than those described for the continuum data.
Therefore, it is likely that the temporal variations are real.

\section{DISCUSSION}

\subsection{Filling Factor of Small Scale \HI\/}

Current theoretical interpretations of the nature of the TSAS 
predict different values for the filling factor of such gas.
\citet{h97} predicts that, essentially independent of the geometry of
the small-scale structures, their volume filling factor would be
approximately 3\%.  Consequently, one would expect a comparable
fraction of lines of sight to show opacity variations.  On the basis
of simulations of the \HI\/ power spectrum, \citet{d00} predicts that
optical depth variations of order 0.2 over transverse scales of
$\sim 100$~AU would be expected approximately 10\% of the time.

Given the extended nature of 3C~138 and high signal to noise of the
data presented in this work, it is possible to observationally
constrain the covering fraction of the TSAS. We use the term `covering
fraction' to indicate that we cannot directly estimate the volume
filling factor, only the fraction of 3C138 covered by TSAS along the
line of sight. For the Epoch II and III data we have estimated the
fraction of pixels exhibiting significant \ion{H}{1} opacity
variations for the 0.9 and 6.2 \kms\/ channels.  We define a
significant opacity change as $\Delta\tau = \mid\tau -
\langle\tau\rangle\mid > 5\sigma_{\tau}$.  In this relation $\tau$ is the
measured opacity at each pixel and $\langle\tau\rangle$ is the mean
opacity over the source. The uncertainty in $\Delta\tau$ is estimated
from adding in quadrature the average rms uncertainty of $\tau$ and the
uncertainty in calculating the mean: $\langle\tau\rangle$. The total
uncertainty is, of course, dominated by the average rms uncertainty in 
$\tau$. Table 3 lists these uncertainties for the 0.9 and 6.2 \kms\/
channels for Epochs II and III. Figure 7 shows histograms of the
number of pixels as a function of $\Delta\tau$ for the 0.9 and 6.2
\kms\/ channels for Epochs II and III.  For each histogram, the bin
size is equal to $1\sigma_{\tau}$ (see Table 3).

With our conservative constraint of $\Delta\tau > 5\sigma_{\tau}$ we find
that the covering fraction of \HI\/ opacity significantly deviating
from the mean for 1999 are: $5\%$ at 0.9 \kms\/ and $11\%$ at 6.2
\kms\/. For 2002 the covering fractions are: $7\%$ at 0.9 \kms\/ and
$\sim 10\%$ at 6.2 \kms\/. From this analysis, we assume a typical
covering fraction of $\sim 10\%$. 

Although we cannot similarly measure the (volume) filling factor it
must necessarily be significantly less than the covering fraction of
$\sim 10\%$. The range of plane-of-sky physical sizes for the clumps
of TSAS range from 5 to 25 AU (for distances between 100 to 500 pc
(\S2); and a typical angular size of 50 mas). Thus, even for aspect
ratios of $\sim 10:1$ a single TSAS enhancement only spans $\sim
10^{-6}$ of the total line of sight.  However, the TSAS is likely
contained within the cold neutral medium (CNM) which itself only
occupies a small fraction of the total line of sight.  Recall that
since 3C~138 is optically visible (\S3.2), most of the CNM is
comprised of \HI\/ along this line of sight, and thus the Aricebo
\HI\/ column density likely traces the bulk of the CNM in this
direction.  If we assume that the average density of the large scale
cold \HI\/ is 56 \cc\/ \citep[i.e. as also assumed by;][based on
pressure arguments]{HTI03b} and use the total cold \HI\/ column
density toward 3C~138 of $6\times 10^{20}$ cm$^{-2}$ (\S3.2), then
{\em all} of the large size scale cold \HI\/ gas toward 3C~138 only
occupies $\sim 3.5$ pc along the line of sight (compared to 500 pc
extent of the Galactic \HI\/ disk in this direction). The line of
sight depth of a single velocity component will be even less
(assuming its density is equal to or higher than 56 \cc\/).  Since
TSAS is observed in all of the velocity components with sufficient
signal to noise to detect it, it seems reasonable to divide the CNM
path length by the number of velocity components (seven).  From these
numbers a lower limit to the filling factor of TSAS in the CNM along
the line of sight to 3C~138 is $\sim 0.1\%$.  Thus, while we emphasize
that we cannot measure this value directly, a TSAS filling factor in
the CNM of $\la 1\%$ is plausible.

\subsection{Pulsar Detection Simulations}

As described in \S 1, contradictory results have been obtained from
attempts to probe the Galactic TSAS with pulsars. These observations
have prompted descriptions of small-scale \ion{H}{1} structure ranging
from potentially a relatively  ``rare phenomenon'' \citep{swhdg03} to
``a general property of the [interstellar medium]'' \citep{fwcm94}.
However, one difficulty with the pulsar observations is that although
they have been sampled on timescales similar to that presented in this
study toward 3C~138 (timescales of a couple of years), as well as 
significantly shorter timescales of a few months, they probe only
discrete lines of sight.  As a simple example, consider a pulsar
moving behind a discrete \ion{H}{1} clump.  With ``poorly chosen''
epochs of observation, the pulsar lines of sight could bracket the
clump and display little or no change in the optical depth.

Our observations cover a continuous range of angular scales from 20 to
300~mas, which includes the scales sampled typically by pulsar
observations. Since the typical size scale of significant \HI\/
opacity variations toward 3C~138 is $\sim 50$ mas (see Figures 4 and
5) we use this value as the fiducial transverse sample scale.  Using
this typical separation we can assess the extent to which multi-epoch
pulsar \ion{H}{1} observations produce ``typical'' optical depth
variations.

A single pulsar ``trial'' is simulated by forming the optical depth
difference $\Delta\tau$ for two points separated by~50~mas along a
cross-cut parallel to the long axis of 3C138 (see Figures~1 and 5).
The angular extent of the long axis is sufficiently long that multiple
``trials'' can be simulated along the cross cut; the trials are made
independent by moving the endpoints at least 20~mas (1 beam width)
along the cross cut.  Moreover, 3C138 is sufficiently wide along its
short axis (SE/NW direction) that multiple, independent (i.e., offset
by at least 20~mas from each other) cross cuts parallel to the long
axis can be formed.  Finally, we conducted pulsar simulation trials at
all three velocities, 0.9, 6.2, and~9.0~\kms\/.  In total, we
were able to simulate 120 pulsar observation trials.

Figure~8 shows the resulting histogram of $\Delta\tau$ values for all
independent pulsar simulations.  It is apparent that the most probable
value for~$\Delta\tau$ is 0, but that there is a tail to large values,
$\Delta\tau \gtrsim 0.3$.  Based on these limited simulations, we
expect that $\Delta\tau < 0.1$ in $\sim90\%$ of pulsar observations.
Our simulated results are in reasonable agreement with the recent
observational results by \citet{swhdg03} and \citet{jkww03} who find
that few pulsars show measurable $\Delta\tau$ changes.  Moreover, the
rapid temporal variations implied by our data toward 3C~138 will
further reduce the probability of detecting variations with pulsar
observations.  Thus, drawing conclusions from pulsar observations
about the fraction of the neutral ISM in small-scale \ion{H}{1}
structures is difficult due to the low probability that the observing
epochs will be exactly matched to the size scale (and timing) of an
appreciable variation in $\Delta\tau$.

\subsection{Significance of Spatial and Time Variations}

\subsubsection{Is the Line of Sight to 3C~138 Special?}

Two of the most notable aspects of the \ion{H}{1} absorption toward
\src\ is the relatively large amplitude of opacity variations observed
and that observed variations have a fairly uniform typical sizescale
of 50 mas ($\sim 25$ AU).  While opacity variations are observed
toward other sources \citep{fgdt98,fg01} in many cases, these opacity
variations are much smaller in amplitude.  To what extent is the line
of sight to \src\ typical of other lines of sight through the
\hbox{ISM}?

Table~4 summarizes the \ion{H}{1} absorption observations
in the literature that have made use of VLBI imaging.  There is a
clear bimodal distribution for the sizes of sources that have been
studied, with only \src\ and \objectname[3C]{3C~147} probing length
scales larger than 100~\hbox{AU}.  In contrast, from Figure 5, it is
clear that for the case of \src\, the typical size scale for significant
changes is $\sim 50$ mas. There are of course uncertainties in
converting between the observed angular scales and the equivalent
linear scales, as in many cases the Galactic \ion{H}{1} rotation curve is of
limited utility in determining the distance to the absorbing gas.
Nonetheless, it is clear that only two sources, \src\ and
\objectname[3C]{3C~147}, are both large enough and have a high enough surface
brightness to make the detection of these opacity variations
significant.

We conclude that the the large amplitude opacity variations that are
observed toward \src\ and \objectname[3C]{3C~147} are likely to be a
selection effect.  Much like the case of the pulsar observations
(\S 4.2), the typical equivalent linear scales probed by
the lines of sight to most sources are too small to have a large 
probability of displaying a significant opacity variation.

\subsubsection{What is the Nature of the Small Scale \HI\/ Gas?}

There are two main explanations for the small scale \HI\/ gas: (1) the
variations are caused by changes in the physical properties of small
parcels of gas; or (2) the variations are merely the manifestation of
random fluctuations contributed by the entire range of spatial
scales in the "red" \HI\/ power spectrum.

In the past, the main argument against a ``physical entity''
interpretation for the small scale structure has been the difficulty
in reconciling the pressure that such ``clumps'' would exert on the
surrounding medium. \citet{h97} has attempted to ameliorate the
pressure problem by suggesting that the temperature of the small scale
features is low compared to the average ISM, and that they have a
filamentary or sheet like structure, so that the density is also lower
than implied by the transverse size. However, we have shown in \S 3.2
that there is little difference between the line widths of the VLBA
spectra and that obtained by \citet{HTI03} using the Arecibo telescope
toward 3C~138. Indeed, the VLBA line widths are slightly {\em wider} than
their Arecibo counterparts (see Table 3).

This comparison seems to exclude the possibility that the small scale
features have significantly lower temperatures.  For example,
\citet{HTI03} estimate from comparison of the Arecibo absorption and
emission spectra that the \HI\/ spin temperatures of the four
strongest 3C~138 velocity components range from 40 to 55 K.  These
spin temperatures translate into thermal line widths of 1.3 to 1.6
\kms\/ (using $\Delta V_{th}= 0.214 T^{0.5}$). The difference between
the observed line width and the thermal width is presumably due to
turbulent motions in the gas. If the spin temperature dropped to $\sim
15$ K in the small scale structures as suggested by \citet{h97}, we
would expect to see a {\em decrease} in the VLBA line widths of $\sim
0.5$ to 0.8 \kms\/ compared to the Arecibo spectrum. Such changes are
well within the sensitivity of the Gaussian fitting for the strong
components, and are not observed. It is possible that a drop in spin
temperature is exactly countered by a large increase in the turbulent
width, but this seems unlikely.  Thus, the absence of line-width
decreases in the VLBA data indicates that the dominant cause of the
opacity variations is likely to be density fluctuations, possibly
accompanied by a small increase ($\lesssim 0.3$ \kms\/) in the thermal
or turbulent contributions to the line widths.  Such an increase in
the thermal or turbulent widths would account for the somewhat greater
widths of the stronger VLBA lines compared to the Arecibo data (Table
2).

\citet{JT01} have carried out a comprehensive survey of the UV lines
of \CI\/ in the directions of 21 early type stars near longitudes of
$100\arcdeg$ and $300\arcdeg$ with distances in the range of 1 to 3.7
kpc. The \CI\/ lines arise from the three fine-structure levels of the
ground electronic state. Based on ratios of the two excited levels, it
is possible to derive the pressures in the interstellar medium. A
small fraction of the sight lines probed by \citet{JT01} show evidence
for significantly enhanced pressures. \citet{Jenkins04} has summarized
these results and suggests that rapid changes in pressure may arise
from the cascade of mechanical energy to small scales from larger
scales. The pressure changes occur over time scales much shorter than
the time required for thermal equilibrium to be
established. \citet{JT01} suggest that one of the motivations for
\citet{h97} to attribute the \HI\/ small scale structure to lower
temperature sheets or filaments viewed edge-on can be eliminated. That
is, the objection to small scale structures based on excessive H$_2$
formation and thus excessive extinction can be avoided with time scale
considerations; H$_2$ molecular formation does not have time to
occur. Thus the overpressure problem, which was the main impetus for
Heiles' suggestions, is solved precisely because these structures are
not in equilibrium and their lifetime is short.  The volume filling
factor of the over pressured ISM, as determined from the \CI\/ lines,
is a few percent at most \citep[$\sim 3.7\%$][]{JT01,Jenkins04},
comparable to the low filling factors of enhanced \HI\/ optical depth
implied by the low covering fraction of such gas toward 3C~138 (see
\S4.1). To summarize, only a small fraction of the total \ion{H}{1}
mass is located in these fluctuations, which is consistent with
various other analysis such as that by Dickey \& Lockman (1990).  These
density fluctuations are almost certainly far from equilibrium and
short-lived.

Currently, only \citet{d00} has attempted the difficult task of
extending the \HI\/ power spectrum down to size scales relevant to the
VLBA opacity variations described here.  Unfortunately, there are a
number of unresolved issues regarding the \citet{d00} extrapolation
that call into question its direct applicability to the current
observational data.  Foremost among these issues is that the observed
\HI\/ power spectrum used in the \citet{d00} study (toward CasA)
only covers one decade in spatial scale, while the extrapolation is
over four decades of spatial scales. As also pointed out by
\citet{d00}, the extrapolation is very sensitive to the assumed
spectral index.  Indeed, the CasA model with a slope of $-2.75\pm
0.25$ \citep{ddg00} predicts peak optical depth variations that are
$\sim 5$ times less than we observe (0.5) toward 3C~138. Moreover,
\citet{ddg00} observed significantly different power law indices for
the line of sight toward Cas~A and one of the two clouds toward
Cygnus~A, suggesting that the power spectrum may depend on direction
\citep[see also][]{green93,dickey01}. 

These observed differences are potentially crucial given that
according to \citet{d00}, a change of just 0.1 in the power law
spectral index would imply a change in the expected opacity variations
on AU scales by a factor of $\sim 2$. Also, if the neutral ISM is
dominated by filamentary or sheet-like structures (as one might expect
if the density fluctuations are transient features arising from
shocks), the effective spectrum of the density fluctuations could be
significantly different than assumed. Further detailed investigation
into these issues on small scales will have to await analysis of the
{\em observed} \HI\/ power spectrum toward 3C~138 down to scales of at
least 10 AU. It is entirely possible that using the appropriate power
spectrum for 3C~138, the \citet{d00} model could produce results that
are in agreement with the level of opacity fluctuations described in
this paper.
 
It is notable that the power law slope determined by \citep{ddg00} and
assumed by \citet{d00}: $-2.75$ is quite shallow compared to that
expected from pure Kolmogorov turbulence of $\sim -3.7$.  Recent
theoretical work by \citet{lazarian00,lazarian04} suggests that if the
velocity, as well as the density structure of the gas are properly
taken into account, the Kolmogorov slope is modified to shallower
values (between $-2.67$ to $-3.4$) depending on whether velocity or
density fluctuations dominate the power spectrum.  The shallowest
values correspond to the velocity dominated case. Indeed,
\citet{lazarian00,lazarian04} suggest that as the velocity averaging
of \HI\/ data is increased, the effect of velocity will be reduced and
the slope should steepen. For \HI\/ observed in emission,
\citet{dickey01} (in the Galaxy) and \citet{Stan01} (in the SMC) do
find evidence of this effect, i.e. the power spectrum slope steepens
as the velocity averaging of the data is increased. However,
\citep{ddg00} do not see evidence of this effect in their \HI\/
absorption data. \citet{dickey01} also find that the degree of
steepening as a function of velocity averaging is much less for the
cooler \HI\/ components in their emission line data compared to the
hotter gas. These observational results are in general agreement with 
\citet{lazarian04} who find that opacity effects can play a 
role in the resulting power spectrum slope, beyond the effects 
of velocity or density dominance. Thus, overall it seems likely 
that all current \HI\/ power spectrum analysis from both 
emission and absorption data \citep[including][]{d00} are consistent 
with a modified Kolmogorov like turbulence.

The fundamental difference between the Jenkins and Deshpande pictures
is that the latter ascribes the small scale fluctuations to ``red''
noise (noise that increases at larger spatial scales) in the \HI\/
power spectrum with contributions from {\em all} spatial scales (i.e.
optical depth changes are statistical and not due to physical changes
in the gas), while the former suggests that a cascade of mechanical
energy creates short lived small physical regions with high pressures.
Our detections of 2-d optical depth enhancements or ``clumps'' (see
Figs. 4, 5 for example) makes the Deshpande picture seem less
likely. Note that this argument would be stronger, even definitive if
the clumps with typical sizes of $\sim 50$ mas were more significantly
larger than the beam size of 20 mas. Future, higher resolution observations and
simulations may help to resolve this issue.  Currently, a paradigm
based on a Kolmogorov like turbulent or mechanical cascade which
produces regions of enhanced pressure does seem favorable based on the
recent theoretical and optical work, in addition to our new \HI\/
optical depth data, although it is quite possible that both phenomena
contribute to the observed opacity differences.

\subsubsection{Significance of Zeeman Non-detections}

The Arecibo Millennium survey provides the highest sensitivity
statistical look at the magnetic field strength in a large sample of
cold neutral medium (CNM) clouds to date
\citep{HT04,Heiles2005,HeilesCr05}, including the direction of
3C~138. For \HI\/ components with either significant \Blos\/
detections (22) or significant upper limits ($\sigma< 10$ \muG\/; 47),
these authors find that (1) the median total magnetic field strength
is $6\pm 1.8$ \muG\/; (2) the median non-thermal contribution to the
linewidth is $\sim 2.8$ \kms\/ (velocity dispersion $\sim 1.2$
\kms\/); (3) there are no strong correlations of \Blos\/ with \HI\/
column density, linewidth, or spin temperature (but the dispersion of
these parameters is also fairly small); and (4) the magnetic and
turbulent energies are in approximate equilibrium \citep[also see
similar results in][]{Myers1995}. Indeed, using the parameters from
(1), (2), and $B=0.4\Delta V_{NT}n^{0.5}$ \citep[obtained by setting
the magnetic and turbulent energies equal, see for
example][]{Crutcher2003,Myers1988} we find that the density is $\sim
30$ \cc\/ in agreement with that expected from pressure equilibrium
arguments \citep{HTI03b}. In the above equation $\Delta V_{NT}$ is the
non-thermal contribution to the FWHM linewidth in \kms\/ and $n$ is
the proton density in \cc\/.  Specifically for the 3C~138 6.2 \kms\/
component, \citet{HT04} derive \Blos\/$=+5.6\pm 1.0$ \muG\/, and
$\Delta V_{NT}=2.1$ \kms\/ (assuming that the thermal contribution has
$T_s\sim 40$ K); this value is also consistent with equipartition
between the magnetic and turbulent energies on large sizescales for
this line of sight if $n\sim 50$ \cc\/.

To avoid potential confusion we note that a similar correlation
between the magnetic field strength and density also exists for dense
molecular clouds with $B\propto \Delta V n^{0.5}$
\citep{Basu2000,Crutcher2003}. However, for self-gravitating clouds it
is unclear whether this relation is the result of equipartition
between the magnetic and turbulent energies (which has also been
observed for this density regime) or depends on the details of
gravity, geometry, and ambipolar diffusion. The observed relation can
be derived equally well from both possible lines of argument;
\citet{HeilesCr05} provide a review of this outstanding question and we
will not address it further here.

As described in \S3.2 and \S4.3.2 the small scale \HI\/ optical depth
enhancements cannot be due to changes in the non-thermal
linewidths. If small scale \HI\/ structures are interpreted as density
enhancements, densities of $\sim 10^5$ \cc\/ are required (see \S1).
Using the non-thermal linewidth of the 6.2 \kms\/ component of $\Delta
V=2.1$ \kms\/, magnetic and turbulent equipartition implies that the
magnetic field strength should be on the order of 250 \muG\/ for
$n\sim 10^5$ \cc\/ -- well within our $3\sigma$ single pixel detection
limit (45 \muG\/; \S3.2). This analysis suggests that magnetic and
turbulent equipartition {\em does not exist} on very small sizescales.
However, since flux freezing almost certainly exists in the cold
neutral medium \citep[see for example][]{HeilesCr05} it is difficult
to imagine a scenario whereby a large change in the density of the
small scale \HI\/ structures -- on the order of two to three orders of
magnitude -- would not result in an appreciable change in the magnetic
field strength. We speculate that the lack of observed increase in
field strength expected from equipartition may be due to an inability
of MHD waves to propagate at these sizescales; MHD waves with
frequencies higher than the ion-neutral collision frequency cannot
propagate \citep[see for example][]{Nakano1998}. For a CNM cloud size
of $\sim 3.5$ pc (\S 4.1), ionization fraction $10^{-4}$
\citep{HeilesCr05}, and Alfv\'en velocity of $\sim 2.1$ \kms\/
(i.e. using the non-thermal linewidth), the cutoff wavelength is on
the order of 10 AU similar to the sizescale of the VLBA \HI\/ features.
Simulations are needed to explore in detail whether this 
or other possible scenarios are plausible.

\subsubsection{Do the \HI\/ Structures Move or Reform?}

Obviously it is impossible with only three epochs of data separated in
time by a few years to answer this question.  From Figure 5, it is
clear that the small scale \HI\/ structures certainly do change on
timescales of $\lesssim 3-4$ years. It is tempting to assign at least
some of the temporal variations visible in Fig. 5 to physical movement
of the \HI\/ ``clumps'' (optical depth enhancements). For example, if
the first clump of the 1995 6.2 \kms\/ data shown in Fig. 5 at a
relative position of $\sim 70$ mas along the cross-cut is the same
feature apparent in the 1999 data at $\sim 150$ mas, and the 2002 data
at $\sim 240$ mas, the ``clump'' is moving with a transverse velocity
of $\sim 20$ \kms\/.  Other similar apparent motions of \HI\/
``clumps'' across the three epochs can also be imagined.  If real,
such motions, which are highly supersonic, could be indicative of
shocks.  Whether the observed enhancements are then due to the clumps
themselves moving or if enhancements reform in new parcels of gas as
the shock front itself moves (i.e. akin to the debate regarding maser
spots) is a matter for further speculation and debate, and is beyond
the scope of this work.  We note that the small increases seen in the
VLBA line widths toward 3C~138 compared to Arecibo data (discussed in
\S4.3.2) provide weak support for a shock model. It is also possible
to reconcile the temporal variations observed toward 3C~138 with the
\citet{d00} picture (where small scale variations are due to
contributions from all spatial scales) if the large scale \HI\/ clouds
are moving with appreciable transverse velocities (A. Deshpande,
private communication).

\section{SUMMARY AND CONCLUSIONS}

We have presented new \ion{H}{1} absorption observations of
\objectname[3C]{3C~138} from~2002 with the \hbox{VLBA} and the
re-analysis of two earlier epochs in~1995 and~1998.  We find
significant spatial and temporal variations in the \ion{H}{1} optical
depth ($\Delta\tau > 0.1$) across the face of \objectname[3C]{3C~138}.
The spatial variations are similar to the levels found by \cite{fg01},
but our signal-to-noise ratio is much improved.

The spatial variations of the \ion{H}{1} optical depth occur on a
typical angular scale of 50~mas and have maximum values of $\sim 0.5$.
Assuming that the absorbing gas is at a distance of no more than
500~pc, the equivalent linear size is $\lesssim 25$~\hbox{AU}. The
total (CNM) \HI\/ column density toward 3C~138 (assuming a spin
temperature of 50 K) is $6\times 10^{20}$ cm$^{-2}$.  Significant time
variations are seen from epoch to epoch, indicating a typical time
scale of no more than a few years.  We have also searched for the
Zeeman effect in the 1999 and 2002 data.  We place a ($3\sigma$) upper
limit of approximately 45~\hbox{$\mu$G} at the 3C~138 continuum peak
and a ($3\sigma$) upper limit of 20 \muG\/ for the averaged \HI\/
profile. We find an average plane of sky covering factor for the small
scale \HI\/ gas of $\sim 10\%$ toward 3C~138; the total volume filling
factor must necessarily be much smaller. A plausible number for the
filling factor of TSAS in the CNM (which itself only
occupies $\sim 1\%$ of the total line of sight) is $1\%$.  We do not
find any evidence for a narrowing of the VLBA \HI\/ line widths
compared to single dish measurements.  This suggest that small scale
variations in the \HI\/ opacities are {\em not} caused by changes in
the spin temperature.

Because of their high velocities, multi-epoch observations of pulsars
would appear to be a way to sample different lines of sight and search
for \ion{H}{1} opacity variations.  Employing this technique, a
variety of pulsar observations have found conflicting results.  We
have simulated a set of pulsar observations by forming the differences
of the \ion{H}{1} opacity across the face of \src\ on an angular scale
comparable to that probed by the pulsar observations.  While we can
find significant opacity variations on a small number of differenced
lines of sight, these occur only with a low probability.
Quantitatively, our simulation is in good 
agreement with the fraction of pulsar observations reporting
significant opacity variations.  We conclude that pulsar observations,
unless they are sampled much more often and over longer durations
than has been the case heretofore, may be of limited utility in
characterizing small-scale opacity variations.

We currently favor the \citet{Jenkins04} interpretation for the nature
of the small scale \HI\/ gas, although the \citet{d00}
interpretation may also be viable if a power spectrum appropriate for
the 3C~138 direction is determined in the future. In the Jenkins
picture, the small scale structure arises from pressure driven density
enhancements due to the cascade of mechanical energy from large to
small scales. The recent theoretical work of \citet{lazarian00}
suggests that the observed power law slopes observed by \citet{ddg00}
and others is in agreement with a revised picture of Kolmogorov
turbulent cascade.  The short lifetime of these structures removes the
need to explain away the TSAS with low temperatures or as merely
``fluctuations'' in the \HI\/ power spectrum. We hesitate to dismiss
the notion that the structures are filamentary or sheet-like, since
evidence for such structures is quite ubiquitous at larger size scales
in recent observations of the ISM from optical (WHAM) to radio
wavelengths (SGPS).

There are a number of future lines of research that would help shed
light on the intriguing results for TSAS presented in this paper. In
particular, a sensitive VLBA \HI\/ monitoring program of the 3C~138
line of sight on short timescales of perhaps a month would help to
determine the timescale for variation. Similar observations toward
other large bright quasars (3C~147 for example) would help to
determine the ubiquity of the \HI\/ variations observed toward
3C~138. Based on our experiences, we recommend that these observations
be carried out with phase referencing to remove the ambiguities
introduced by self-fringing on the target source. A time variability
monitoring program for the optical fine structure lines would also be
interesting.  A power spectrum from large (100pc) to small (10 AU)
sizescales should be constructed for the \HI\/ gas toward 3C~138 using
existing VLA, MERLIN, and VLBA data, and compared in detail with the
predictions of the Deshpande model.  Theoretical
constraints/predictions for the timescale and level of density
contrast possible from a Kolmogorov-like cascade would also provide
useful comparisons with the observational data. It seems likely that
such a cascade would require a high frequency cutoff (corresponding to
small spatial scales) in order to produce a strong density contrast
and hence the TSAS -- such a cutoff may also explain the typical
sizescale of TSAS observed in the data presented here of $\sim 50$ mas
($\sim 25$ AU). For example, some form of MHD waves may provide such a
small scale cutoff, since these waves cannot propogate at frequencies
higher than the ion-neutral collision rate.

\bigskip

\acknowledgments 

We thank A. A. Deshpande, R. J. Reynolds, and S. Stanimirovi{\'c} for
valuable comments and discussions on the manuscript.  We would also
like to thank the anonymous referee for helping us to improve the
clarity of the manuscript.  Basic research in radio astronomy at the
Naval Research Laboratory is supported by the Office of Naval
Research.

\begin{figure}[h!]
\epsscale{1.0}
\plotone{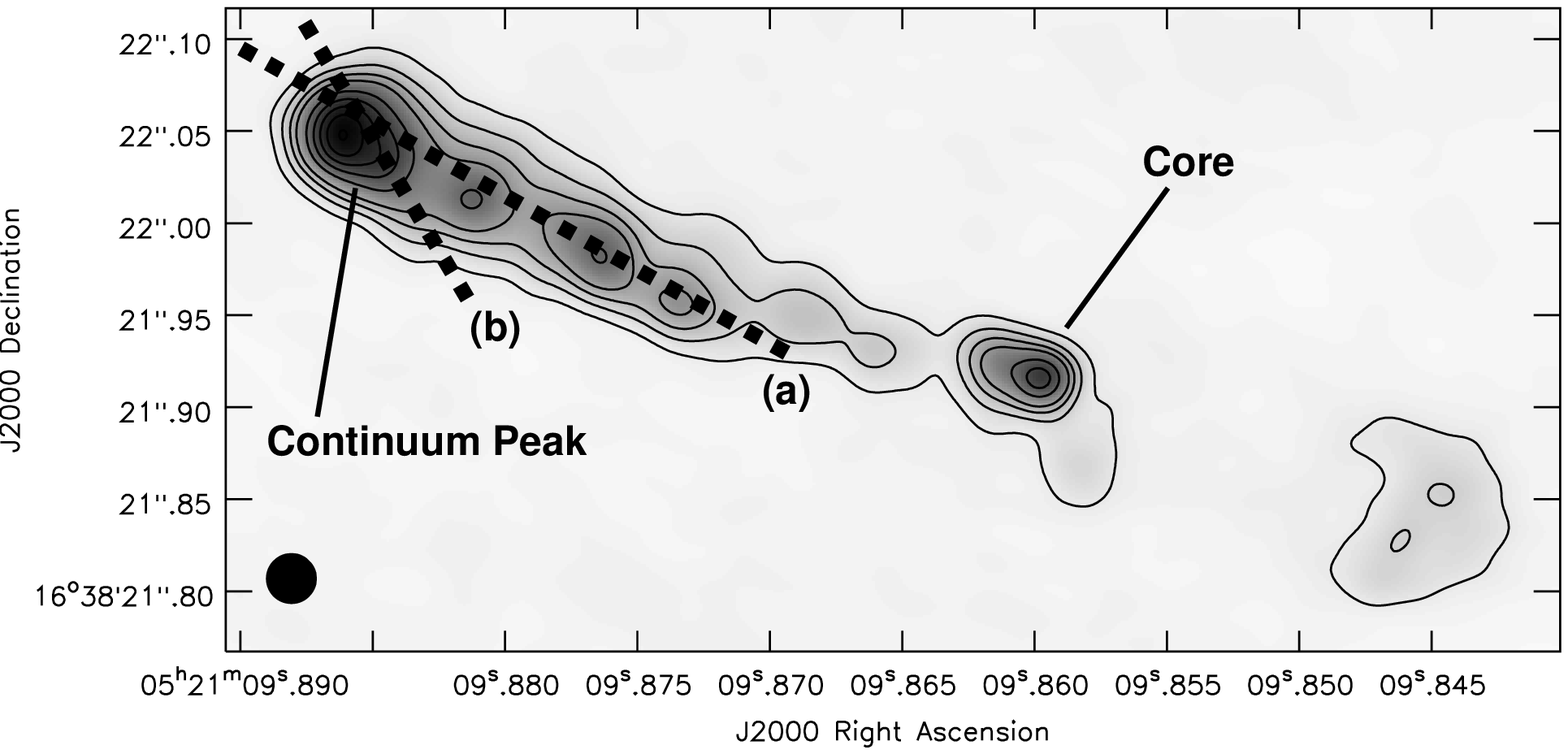}

\caption[]{VLBA 21cm continuum image of 3C~138 with 20 mas resolution
from Epoch III (2002).  The black contours are at 10, 30, 60, 100,
200, 300, 400, 500, and 600 \mjb\/. The peak flux density is 607 \mjb\/ and 
the rms noise is 0.95 \mjb\/. The 60 \mjb\/ contour level is used for a 
continuum cutoff in subsequent optical depth discussions. The cross-cuts 
labeled (a) and (b) are described in \S 3.2.}

\end{figure}

\begin{figure}[h!]
\epsscale{0.5}
\plotone{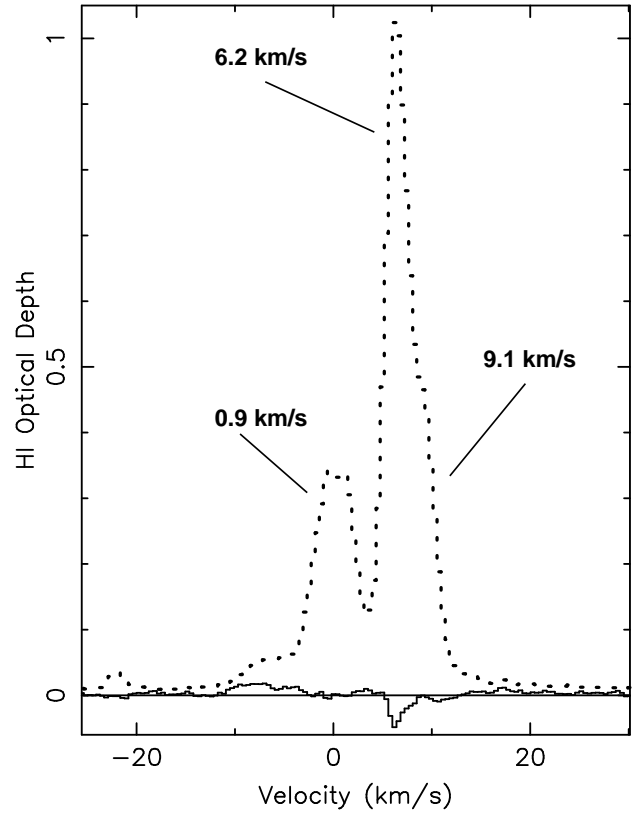}

\caption[]{Average optical depth profile from the Epoch III (2002)
data ({\em dotted}).  The three velocity components discussed in the
text are labeled for reference. The solid line shows the difference
between the average 2002 and average 1999 optical depth profiles 
(i.e. $\tau_{\rm Avg{2002}}-\tau_{\rm Avg1999}$)}.
\end{figure}

\begin{figure}[h!]
\epsscale{1.0}
\plotone{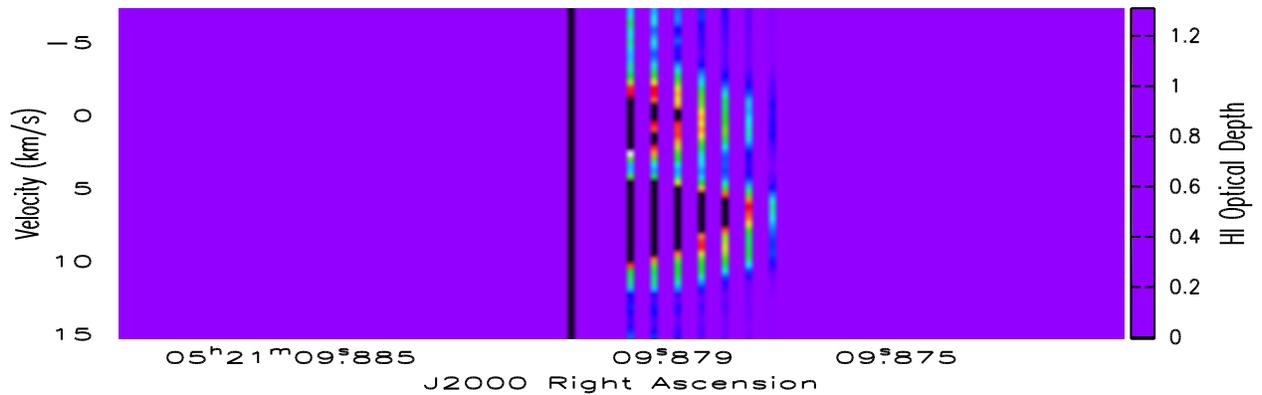}

\caption[]{Position-Velocity diagram from the Epoch III (2002) data 
along the cross-cut labeled (a) in Figure 1.}
\end{figure}

\begin{figure}[h!]
\includegraphics[angle=90,scale=0.8]{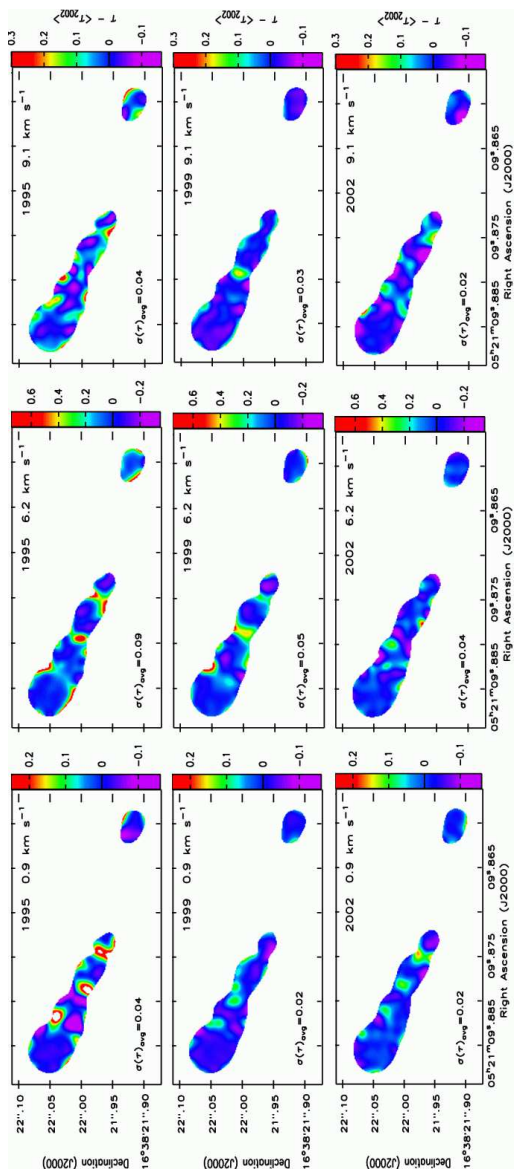}

\caption[]{Optical depth channel maps minus the average value of the
2002 optical depth ($\tau - <\tau_{2002}>$) with velocity increasing
to the right and time increasing down. The color scale for a given
velocity is constant for each epoch. The color scale has been adjusted
so that the dark blue color is centered about zero. The 2002 average
optical depth values that were subtracted are: 0.34 at 0.9 \kms\/,
1.02 at 6.2 \kms\/, and 0.47 at 9.1 \kms\/ (Table~2). For reference
the {\em average} value of the $1 \sigma$ uncertainty in $\tau$
($\sigma(\tau)_{avg}$) is indicated in the lower left hand
corner. Note that while the 1995 data seem to have the largest
variations, these data are also noisier by a fact of $\sim 2$ than the
1999 or 2002 data. Thus, it is very important to take into account the
noise estimates when interpreting this Figure. For further discussion
of the uncertainties see \S 3.4 and Table~2.  }

\end{figure}

\begin{figure}[h!]
\epsscale{1.0}
\plotone{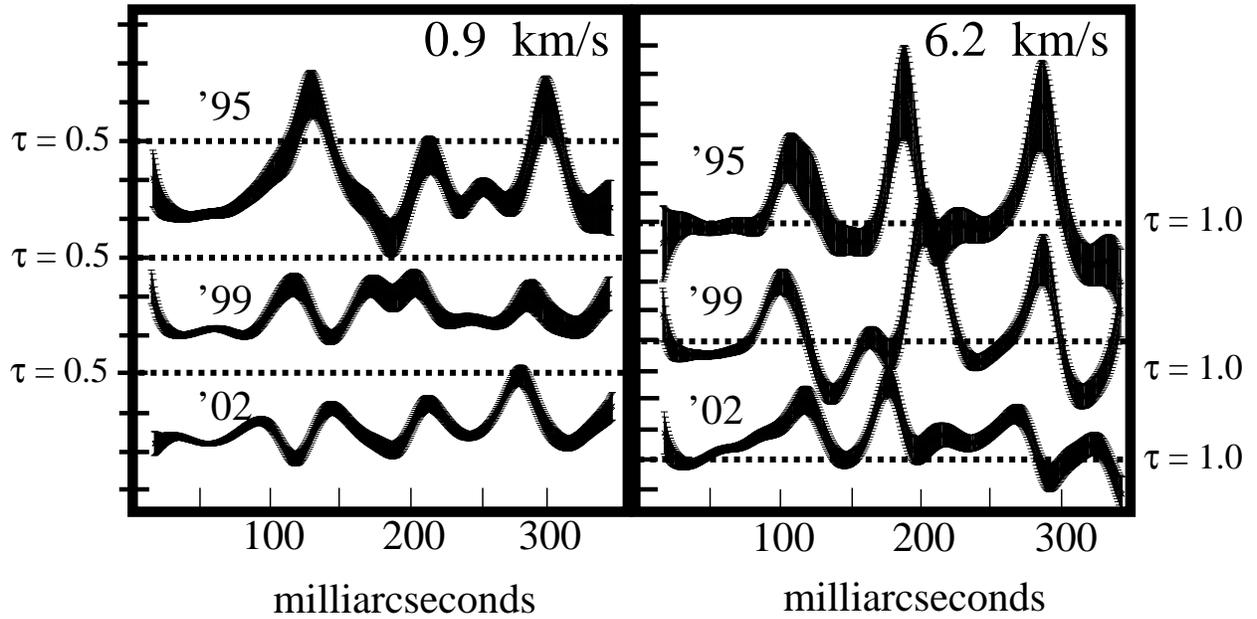}

\caption[]{Optical depth cross-cuts at 0.9 and 6.2 \kms\/ along the
position marked (a) in Figure 1 for all three epochs. The vertical tic
mark interval is 0.1 and the width of the shaded line is equal to $\pm
1\sigma$, where $\sigma$ is the uncertainty in the measurement of
$\tau$.  This figure shows that the typical linear distance between
significant \HI\/ variations is 50~mas.}

\end{figure}
\begin{figure}[h!]
\epsscale{0.5}
\plotone{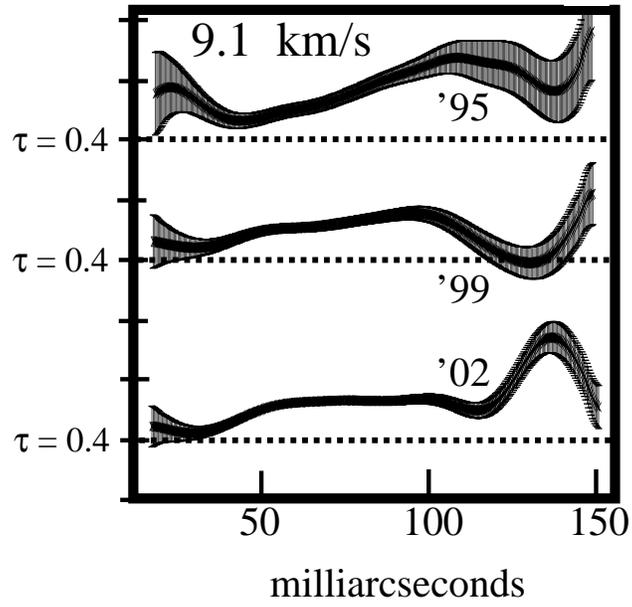}

\caption[]{Optical depth cross-cuts at 9.1 \kms\/ along the position
marked (b) in Figure 1 for all three epochs.  The vertical
tic mark interval is 0.05 and the width of the shaded
line is equal to $\pm 1\sigma$, where $\sigma$ is the uncertainty in
the measurement of $\tau$.  This Figure is an example of of a
cross-cut where no significant variations are observed.}

\end{figure}

\begin{figure}[h!]
\epsscale{0.95}
\plotone{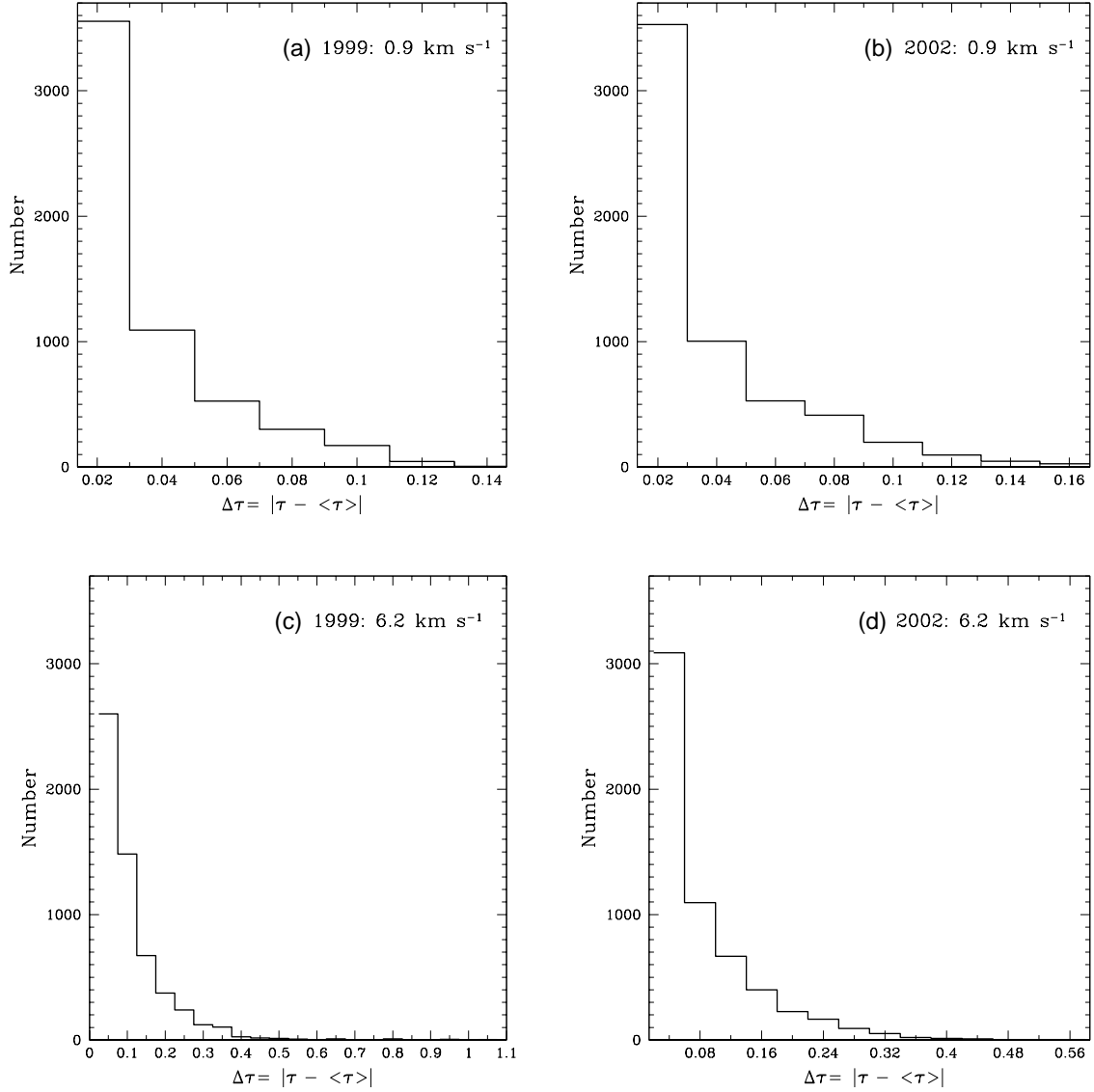}

\caption[]{Plot of the number of pixels in the (a) Epoch II 0.9 \kms\/, 
(b) Epoch III 0.9 \kms\/, (c) Epoch II 6.2 \kms\/, and (d) Epoch III 
6.2 \kms\/ opacity images as a function of $\Delta\tau = \mid\tau -
  \langle\tau\rangle\mid$.  The bin size is equal to the $1\sigma$
  uncertainty in $\Delta\tau$ (see Table 3). The total number of
  pixels sampled is 5693 and 5837 for the Epoch II and Epoch III data,
  respectively. The percentage of pixels with $\Delta\tau\gtrsim 5\sigma$ is 
$5\%, 7\%, 11\%$, and $10\%$ for panels (a)-(d) respectively.}

\end{figure}

\begin{figure}[h!]
\epsscale{0.5}
\plotone{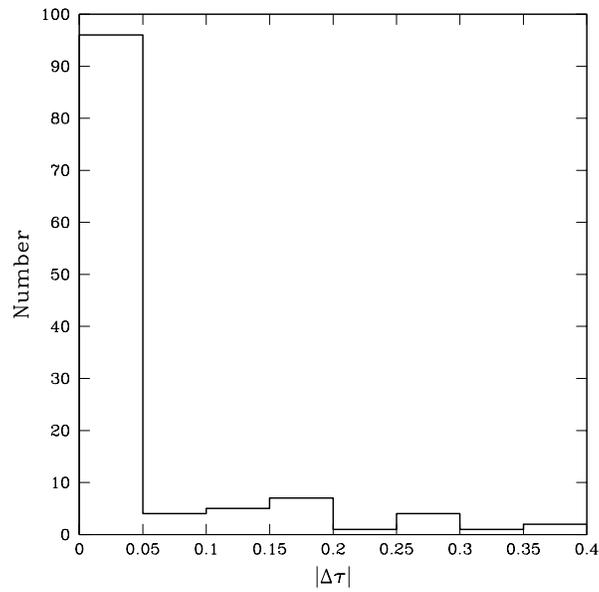}

\caption[]{Simulation of the $\Delta\tau$ variations that would be 
observed by 120 pulsar observations with epochs separated by the 
equivalent of 50 mas (typical size scale of \HI\/ opacity variations, see 
Fig. 4). The beginning and end points of the observations 
are randomly selected from the cross-cut labeled (a) in Figure 1 as well 
as two parallel positions separated by a beam width.}

\end{figure}

\newpage

\begin{deluxetable}{lccc}
\small
\tablewidth{45pc}
\tablecaption{Observational Parameters\label{tab1}}
\tablecolumns{4}
\tablehead{
\colhead{Parameter} & \colhead{Epoch I} &\colhead{Epoch II} & \colhead{Epoch III}}
\startdata
Date & 10 Sept 1995 & 22 Dec 1999 & 12 May 2002 \\
Number of IFs &  4 & 4 & 4 \\
Bandwidth per IF (MHz) & 0.25 & 0.5 & 0.5 \\
Spectral channels & 128 & 256 & 256 \\
Channel separation (\kms\/) & 0.41 & 0.41 & 0.41 \\
Velocity resolution$^a$ (\kms\/) & 0.82 & 0.82 & 0.82 \\
Clean Beam$^d$ (mas) & 18.7 x 13.6  ($-$0.7) & 17.9 x 10.3 ($-$6.5) & 19.7 x 14.5 (3.5)\\
Continuum Peak (\mjb\/) & 610 & 600 & 607 \\
Continuum rms noise (\mjb\/) & 1.1  & 0.88  & 0.95  \\
Spectral line rms noise (\mjb\/) & 3.0  & 1.8  & 1.6  \\
\cutinhead{General Parameters for 3C~138}
Position$^b$ (J2000) &  $05^{\rm h}21^{\rm m}09\fs 904$, & $+16\arcdeg 38\arcmin 22\farcs116$ \\
Position$^c$ (Galactic) & $+187.4\arcdeg$, &  $-11.34\arcdeg$  & \\ 
Redshift  & 0.759 & & \\
\enddata
\tablenotetext{a} {After Hanning smoothing during imaging process.}
\tablenotetext{b} {Right Ascension, Declination}
\tablenotetext{c} {Longitude, Latitude}
\tablenotetext{d} {These are the clean beams before convolution to 20 mas. 
Number in parenthesis is the position angle of the beam in degrees.
All subsequent parameters are for the convolved 20 mas resolution images.}
\end{deluxetable}

\begin{deluxetable}{ccccccc}
\small
\tablewidth{40pc}
\tablecaption{Results From \HI\/ Gaussian Analysis}
\tablecolumns{7}
\tablehead{
\colhead{ } & 2002 VLBA$^a$ &\colhead{}  & \colhead{~~~~~~~~~ } & \colhead{ } 
& 2000 Arecibo$^{b,c}$ 
& \colhead{ } \\
\colhead{$V_{LSR}$ (\kms\/)} & \colhead{$\Delta V$ (\kms\/)} & \colhead{$\tau^d$} & 
\colhead{~~~~~~~~~ } & \colhead{$V_{LSR}$ (\kms\/)} & \colhead{$\Delta V$ (\kms\/)} 
&\colhead{$\tau^e$}}

\startdata
$-$21.5  & $2.8\pm 0.8$ & 0.02 & & $-$21.5 & $3.5\pm 0.2$ & 0.04\\
$-$7.3   & $2.1\pm 0.9$ & 0.02 &  &  - & - & - \\
$-$0.4 & $3.3\pm 0.1$ & 0.27 & & $-$0.5 & $2.9\pm 0.1$ & 0.25\\
+1.7  & $2.1\pm 0.1$ & 0.13 &  & +1.6 & $1.8\pm 0.1$ & 0.18\\
+1.9 & $19.0\pm 1.0$ & 0.07 &  & +1.8 & $14.6\pm 0.5$ & 0.06\\
+6.4 & $2.56\pm 0.02$ & 0.93 &  & +6.4 & $2.30\pm 0.02$ & 1.05\\
+9.1 & $2.87\pm 0.05$ & 0.37 &  & +9.1 & $2.81\pm 0.06$ & 0.46\\
\enddata
\tablenotetext{a} {Position from continuum peak}
\tablenotetext{b} {Arecibo data from \citet{HTI03}}
\tablenotetext{c} {Effective resolution is the source size of $\sim 800$ mas}
\tablenotetext{c} {Average error on VLBA $\tau$ fit is 0.007}
\tablenotetext{c} {Average error on Arecibo $\tau$ fit is 0.006}
\end{deluxetable}

\begin{deluxetable}{lccccc}
\small
\tablewidth{38pc}
\tablecaption{Uncertainties in \HI\/ Optical Depths\label{tab1}}
\tablecolumns{6}
\tablehead{
\colhead{Epoch}& \colhead{Velocity (\kms\/)} 
& \colhead{ $\langle\tau\rangle$} & \colhead{Min. $\tau$ error$^a$} & 
\colhead{Max. $\tau$ error$^a$ } & \colhead{Avg. $\tau$ error ($\sigma_{\tau}$)}}
\startdata
1995 & 0.9  & 0.34 & 0.007 & 0.09 & 0.04\\
& 6.2  & 1.11 & 0.014 & 0.40  & 0.09\\
& 9.1 & 0.50 & 0.009 & 0.11 & 0.04\\
\hline
1999 & 0.9 & 0.34 & 0.004 & 0.05  & 0.02\\
& 6.2 & 1.07 & 0.009 & 0.28 & 0.05 \\ 
& 9.1 & 0.47  & 0.005 & 0.06 & 0.03\\
\hline
2002 & 0.9 & 0.34 & 0.003 & 0.04 &  0.02 \\
& 6.2 & 1.02 & 0.007 & 0.13 &  0.04 \\
& 9.1 &  0.47 & 0.004 & 0.05 & 0.02\\
\enddata
\tablenotetext{a} {The uncertainty in $\tau$ is inversely proportional to the 
continuum strength. i.e. minima in $\tau$ error occur near continuum peaks and 
maxima in $\tau$ error occur where continuum is weak. }
\end{deluxetable}

\begin{deluxetable}{lcccccccc}
\tablecaption{VLBI \ion{H}{1} Absorption Measurements toward Compact
        Extragalactic Sources\label{tab:vlbi}}
\tablewidth{0pc}
\tabletypesize{\small}
\tablecolumns{9}
\tablehead{
 \colhead{} & 
 \colhead{Galactic} & 
 \colhead{\ion{H}{1} Gas} &
 \colhead{Flux}     &
 \colhead{Angular } & 
 \colhead{Linear} &
 \colhead{Largest$^a$ } & 
 \colhead{Largest } & 
 \colhead{} \\
 \colhead{Name} & 
 \colhead{latitude} & 
 \colhead{Distance} & 
 \colhead{Density}  &
 \colhead{Resolution} & 
 \colhead{Resolution} &
 \colhead{Angular Scale} & 
 \colhead{Linear Scale} & 
 \colhead{Ref.} \\
 \colhead{} & 
 \colhead{(\arcdeg)} &
 \colhead{(pc)} & 
 \colhead{(Jy)} &
 \colhead{(mas)} & 
 \colhead{(AU)} &
 \colhead{(mas)} & 
 \colhead{(AU)} & 
 \colhead{}}

\startdata
\objectname[3C]{3C~119}    & $-4$  & $< 500$ &    8.6 & 10 & 5 & 50 & 25 & 2 \\
\objectname[3C]{3C~138}    & $-11$ & $< 500$ &    8.5 & 20 & 10 & 400 & 200 & 1\\
\objectname[3C]{3C~147}    & $+10$ & 500--1000 & 22.5 & 10 & 10 & 190 & 200 & 2\\

\objectname[]{B0404$+$768} & $+18$ & $< 320$ &    5.8 & 10 &  3 & 100 &  30 & 1\\
\objectname[]{B0831$+$557} & $+36$ & 200     &    8.8 & 20 & 4 & 160 & 30 & 2 \\
\objectname[]{B2255$+$416} & $-16$ & $< 360$ &    2.1 & 10 &  4 &  40 &  16 & 1\\
\objectname[]{B2352$+$495} & $-12$ & $\approx 1300$ & 2.4 & 10 & 13 & 50 & 70 & 2\\
\enddata

\tablerefs{(1)~\cite{fgdt98}; (2)~\cite{fg01}}
\tablenotetext{a} {Sizescale of most extended feature or distance between compact 
features, whichever is largest.}
\end{deluxetable}

\end{document}